# Thermal radiative and thermodynamic properties of solid and liquid uranium and plutonium carbides in the visible-near infrared range

Anatoliy I Fisenko, Vladimir F Lemberg

*ONCFEC Inc., 250 Lake Street, Suite 909, St. Catharines, Ontario L2R 5Z4, Canada*
*E-mail: afisenko@oncfec.com*

**Abstract**

The knowledge of thermal radiative and thermodynamic properties of uranium and plutonium carbides under extreme conditions is essential for designing a new metallic fuel materials for next generation of a nuclear reactor. The present work is devoted to the study of the thermal radiative and thermodynamic properties of liquid and solid uranium and plutonium carbides at their melting/freezing temperatures. The Stefan-Boltzmann law, total energy density, number density of photons, Helmholtz free energy density, internal energy density, enthalpy density, entropy density, heat capacity at constant volume, pressure, and normal total emissivity are calculated using experimental data for the frequency dependence of the normal spectral emissivity of liquid and solid uranium and plutonium carbides in the visible-near infrared range. It is shown that the thermal radiative and thermodynamic functions of uranium carbide have a slight difference during liquid-to-solid transition. Unlike UC, such a difference between these functions have not been established for plutonium carbide. The calculated values for the normal total emissivity of uranium and plutonium carbides at their melting temperatures is in good agreement with experimental data. The obtained results allow to calculate the thermal radiative and thermodynamic properties of liquid and solid uranium and plutonium carbides for any size of samples. Based on the model of Hagen-Rubens and the Wiedemann-Franz law, a new method to determine the thermal conductivity of metals and carbides at the melting points is proposed.

**Keywords** Solid/liquid uranium and plutonium carbides; Melting/freezing temperature; Stefan–Boltzmann's law; Normal total emissivity; Free energy; Thermal conductivity

## 1 Introduction

It is now recognized that uranium and plutonium carbides are candidates for use them as metallic fuel materials for the Generation IV nuclear plant reactors.[1,2] These nuclear fuel materials have a higher thermal conductivity and fissile metal density in comparison to the uranium and plutonium oxide fuels. Due to their high melting points, they are capable of withstanding very high temperatures during normal and accidental operation of a reactor.

Currently, extensive research is being conducted to study the thermodynamic properties of the uranium and plutonium carbides at high temperatures.[3-7] These properties include: a) thermal conductivity; b) heat capacity at constant pressure; c) coefficient of thermal expansion; and etc.

It is essential to note that little data on the thermal radiative properties of UC and PuC during liquid-to-solid transition can be obtained in the literature. In [8], the normal spectral emissivity ($\varepsilon_{0.65\,\mu\mathrm{m}}$) as well as the total normal emissivity ($\varepsilon_t$) of the uranium mono-and dicarbide have been systematically studied between room temperature and 4200 K. The melting temperature was reported to be (2780 ± 15) K. The normal spectral emissivity ($\varepsilon_{0.65\,\mu\mathrm{m}}$) measurements on UC in the temperature range from 1200 K to 2240 K have been made in [9]. Using experimental data, the analytical equations representing the mean values of $\varepsilon_{0.65\,\mu\mathrm{m}}$ and $\varepsilon_t$ for UC at high temperatures are presented in [6]. The average error of approximations are: a) ± 21% for $\varepsilon_{0.65\,\mu\mathrm{m}}$ and b) ± 40% for the normal total emissivity $\varepsilon_t$.

A detailed study of the wavelength dependence of the normal spectral emissivity of solid and liquid uranium and plutonium carbides in the visible-near infrared range at the melting points (2780 ± 15) K and (1900 ± 20) K was conducted in [10]. A slight deference between the normal spectral emissivity in solid and liquid phases was observed for uranium carbide, whereas a similar effect was insensitive to the phase change on melting for plutonium carbides.

At present, there is considerable interest to obtain new data on the thermal radiative and thermodynamic properties of liquid and solid uranium and plutonium carbides at the melting temperatures.

In [11-14], it was shown that the knowledge of the frequency dependence of the normal spectral emissivity $\varepsilon(\nu,T)$ at high temperatures allow to obtain the temperature dependences of the radiative and thermodynamic properties of a real-body. The performance of this method was demonstrated on the thermal radiation of: a) $ZrB_2$-SiC-based ultra- high temperature ceramic [14]; b) zirconium, hafnium and titanium carbides [13]; and other materials [11,12].

The present research focuses on the study of the thermal radiative and thermodynamic properties of liquid and solid uranium and plutonium carbides in the frequency range $0.326\,\mathrm{PHz} \le \nu \le 0.545\,\mathrm{PHz}$ at their melting/freezing temperatures. Using the analytical expression for the normal spectral emissivity on the wavelength [10], the Stefan-Boltzmann law, total energy density, normal total emissivity $\varepsilon(T)$, Helmholtz free energy, entropy, heat capacity at constant volume, pressure, enthalpy, and internal energy are calculated in the liquid and solid phases. It is shown that the slight differences between the calculated values of the thermal radiative and thermodynamic functions of solid/liquid UC are observed. However, the thermal radiative and thermodynamics functions of PuC do not change during solid–to-liquid transition. A method for determining the thermal conductivity of carbides and metals at their melting/freezing temperatures is proposed. This method is based on the Hagen-Rubens model and Wiedemann–Franz law.

## 2. Theory. General relationships.

### *2.1 Thermal radiative properties of a real-body*

The spectral energy density function of thermal radiation of a real-body having emitted continuous spectra can be presented in the form,

$$I(\nu,T) = \varepsilon(\nu,T) I^{\mathrm{P}}(\nu,T), \tag{1}$$

where $\varepsilon(\nu,T)$ is the spectral emissivity and $I^{\mathrm{P}}(\nu,T)$ at temperature $T$ is given by the Planck law [15]:

$$I^{\mathrm{P}}(\nu,T) = \frac{8\pi h}{c^3} \frac{\nu^3}{e^{\frac{h\nu}{k_{\mathrm{B}}T}} - 1} \quad . \tag{2}$$

The total energy density of a real body $I(\nu_1,\nu_2,T)$ can be defined as

$$I(\nu_1,\nu_2,T) = \int_{\nu_1}^{\nu_2} \varepsilon(\nu,T) I^{\mathrm{P}}(\nu,T) d\nu \quad . \tag{3}$$

Using the relationship between the total energy density and the total radiation power per unit area $I^{\mathrm{SB}} = \frac{c}{4} I$ , the Stefan-Boltzmann law in the finite frequency range takes the form

$$I^{\mathrm{SB}} = \frac{c}{4}\int_{\nu_1}^{\nu_2} \varepsilon(\nu,T) I^{\mathrm{P}}(\nu,T) d\nu \,. \tag{4}$$

Using Eq. 1, the total emissivity can be presented in the form,

$$\varepsilon(\nu_1,\nu_2,T) = \frac{I(\nu_1,\nu_2,T)}{I^{\mathrm{BB}}(\nu_1,\nu_2,T)} \,, \tag{5}$$

where $I^{\mathrm{BB}}(\nu_1,\nu_2,T)$ is the total energy density of black-body radiation. The analytical expression for $I^{\mathrm{BB}}(\nu_1,\nu_2,T)$ is determined by the following expression [16]:

$$I^{\mathrm{BB}}(\nu_1,\nu_2,T) = \frac{48\pi(k_{\mathrm{B}}T)^4}{c^3h^3}[P_3(x_1) - P_3(x_2)] \,. \tag{6}$$

Here $x = \frac{h\nu}{k_{\mathrm{B}}T}$ and $P_3(x)$ has the form,

$$P_3(x) = \sum_{s=0}^{3} \frac{(x)^s}{s!} \mathrm{Li}_{4-s}(e^{-x}) \,, \tag{7}$$

where

$$\mathrm{Li}_{4-s}(e^{-x}) = \sum_{k=1}^{\infty} \frac{e^{-kx}}{k^{4-s}} \quad , \left|e^{-kx}\right| < 1 \tag{8}$$

is the polylogarithm functions [17].

The number density of photons of a real-body radiation is presented by the formula [15]:

$$n = \frac{8\pi}{c^3}\int_{\nu_1}^{\nu_2} \frac{\varepsilon(\nu,T)\nu^2}{e^{\frac{h\nu}{k_{\mathrm{B}}T}} - 1} d\nu \,. \tag{9}$$

2.2 *Thermodynamics of thermal radiation of a real-body*

According to [15], the Helmholtz free energy density can be represented in the form,

$$f(\nu_1,\nu_2,T) = \frac{8\pi k_{\mathrm{B}}}{c^3}\int_{\nu_1}^{\nu_2} \nu^2 \varepsilon(\nu,T) \ln\left((1 - e^{-\frac{h\nu}{k_{\mathrm{B}}T}}\right) d\nu \,. \tag{10}$$

Here $\varepsilon(\nu,T)$ is the spectral emissivity of a real-body.

Using Eqs. 3 and 10, the thermodynamic functions of thermal radiation of a real-body can be presented by the following formulas:

1. Entropy density $s$

$$s = -\frac{\partial f}{\partial T} \quad ; \tag{11}$$

2. Heat capacity at constant volume per unit volume $c_V$

$$c_V = \left( \frac{\partial I(v_1, v_2, T)}{\partial T} \right)_V ; \tag{12}$$

3. Pressure $p$

$$p = -f \; ; \tag{13}$$

4. Internal energy density $u$

$$u = f + Ts ; \tag{14}$$

5. Enthalpy density $h$

$$h = u + p ; \tag{15}$$

6. Gibbs free energy density $g$

$$g = h - Ts \; ; \tag{16}$$

7. Chemical potential density $\mu$

$$\mu = \left( \frac{\partial g}{\partial n} \right)_{T,V} . \tag{17}$$

## 3. Calculations

### *3.1 Thermal radiative properties of materials at melting/freezing temperatures*

There are several classes of materials for which the normal spectral emissivity in the solid and liquid phases can be represented in the form,

$$\varepsilon(\nu, T) = \tilde{a} - \tilde{b}\,\nu^{-1} + \tilde{c}\,\nu^{-2} , \tag{18}$$

where $\tilde{a}, \tilde{b}$, and $\tilde{c}$ are constants. These materials include: a) Carbides - ZrC, UC, and PuC [10,18]; b) Noble metals - Cu, Ag, and Au [19,20]; and c) Metals - iron, cobalt, and nickel [21].

Substituting Eq. 18 in Eq. 3 and after the integration over a finite range of frequencies, the total energy density at melting point takes the following form:

$$I(T)=\frac{48\pi k_B{}^4T^4\tilde{a}}{c^3h^3}(P_3(x_2)-P_3(x_1))[1-\frac{h\tilde{b}}{3k_BT\tilde{a}}\frac{(P_2(x_2)-P_2(x_1))}{(P_3(x_2)-P_3(x_1))}+\frac{h^2\tilde{c}}{6k_B{}^2T^2\tilde{a}}\frac{(P_1(x_2)-P_1(x_1))}{(P_3(x_2)-P_3(x_1))}]\ . \quad (19)$$

In accordance with Eq. 4, the total radiation power per unit area can be presented in the form,

$$I^{\mathrm{SB}}(T)=\frac{12\pi k_B{}^4T^4\tilde{a}}{c^2h^3}(P_3(x_2)-P_3(x_1))[1-\frac{h\tilde{b}}{3k_BT\tilde{a}}\frac{(P_2(x_2)-P_2(x_1))}{(P_3(x_2)-P_3(x_1))}+\frac{h^2\tilde{c}}{6k_B{}^2T^2\tilde{a}}\frac{(P_1(x_2)-P_1(x_1))}{(P_3(x_2)-P_3(x_1))}]. \quad (20)$$

Using Eqs. 5, 6 and 18, the normal total emissivity is defined by the following expression:

$$\varepsilon(T)=\tilde{a}[1-\frac{h\tilde{b}}{3k_BT\tilde{a}}\frac{(P_2(x_2)-P_2(x_1))}{(P_3(x_2)-P_3(x_1))}+\frac{h^2\tilde{c}}{6k_B{}^2T^2\tilde{a}}\frac{(P_1(x_2)-P_1(x_1))}{(P_3(x_2)-P_3(x_1))}]\,. \quad (21)$$

Number density of photons, according to Eqs. 9 and 18, is presented by the formula:

$$n=\frac{16\pi k_B{}^3T^3\tilde{a}}{c^3h^3}(P_2(x_2)-P_2(x_1))[1-\frac{h\tilde{b}}{2k_BT\tilde{a}}\frac{(P_1(x_2)-P_1(x_1))}{(P_2(x_2)-P_2(x_1))}+\frac{h^2\tilde{c}}{2k_B{}^2T^2\tilde{a}}\frac{(P_0(x_2)-P_0(x_1))}{(P_2(x_2)-P_2(x_1))}]\ . \quad (22)$$

*3.2 Thermodynamic functions of thermal radiation of materials at their melting points***.**

Using Eq. 19 and after computing the integral in Eq. 10 over the finite frequency range, the general expressions for the thermodynamic functions of thermal radiation of materials during liquid-to-solid transition can be expressed as follows:

a) Helmholtz free energy density *f*

$$f=\tilde{a}f_0+\tilde{b}f_{-1}+\tilde{c}f_{-2}\ , \quad (23)$$

where

$$f_0=-\frac{16\pi k_B{}^4}{c^3h^3}T^4\left\{[P_3(x_1)-P_3(x_2)]-\frac{1}{6}\left(x_1^3\mathrm{Li}_1(e^{-x_1})-x_2^3\mathrm{Li}_1(e^{-x_2})\right)\right\}; \quad (24)$$

$$f_{-1}=-\frac{8\pi k_B{}^3}{c^3h^2}T^3\left\{[P_2(x_1)-P_2(x_2)]-\frac{1}{2}\left(x_1^2\mathrm{Li}_1(e^{-x_1})-x_2^2\mathrm{Li}_1(e^{-x_2})\right)\right\}; \quad (25)$$

$$f_{-2}=-\frac{8\pi k_B{}^2}{c^3h}T^2\left\{[P_1(x_1)-P_1(x_2)]-\left(x_1\mathrm{Li}_1(e^{-x_1})-x_2\mathrm{Li}_1(e^{-x_2})\right)\right\}. \quad (26)$$

b) Entropy density *s*

$$s=\tilde{a}s_0+\tilde{b}s_{-1}+\tilde{c}s_{-2}\,, \quad (27)$$

where

$$s_0 = \frac{64\pi k_B{}^4}{c^3h^3}T^3\left\{\left[P_3(x_1)-P_3(x_2)\right]-\frac{1}{24}\left(x_1^3\mathrm{Li}_1(e^{-x_1})-x_2^3\mathrm{Li}_1(e^{-x_2})\right)\right\}; \tag{28}$$

$$s_{-1} = \frac{24\pi k_B{}^3}{c^3h^2}T^2\left\{\left[P_2(x_1)-P_2(x_2)\right]-\frac{1}{6}\left(x_1^2\mathrm{Li}_1(e^{-x_1})-x_1^2\mathrm{Li}_1(e^{-x_2})\right)\right\}; \tag{29}$$

$$s_{-2} = \frac{16\pi k_B{}^2}{c^3h}T\left\{\left[P_1(x_1)-P_1(x_2)\right]-\frac{1}{2}\left(x_1\mathrm{Li}_1(e^{-x_1})-x_2\mathrm{Li}_1(e^{-x_2})\right)\right\}. \tag{30}$$

c) Heat capacity at constant volume per volume $c_V$

$$c_V = \tilde{a}c_{V0} + \tilde{b}c_{V-1} + \tilde{c}c_{V-2}, \tag{31}$$

where

$$c_{V0} = \frac{192\pi k_B{}^4}{c^3h^3}T^3\left\{\left[P_3(x_1)-P_3(x_2)\right]+\frac{1}{24}\left(x_1^4\mathrm{Li}_0(e^{-x_1})-x_2^4\mathrm{Li}_0(e^{-x_2})\right)\right\}; \tag{32}$$

$$c_{V-1} = \frac{48\pi k_B{}^3}{c^3h^2}T^2\left\{\left[P_2(x_1)-P_2(x_2)\right]+\frac{1}{6}\left(x_1^3\mathrm{Li}_0(e^{-x_1})-x_2^3\mathrm{Li}_0(e^{-x_2})\right)\right\}; \tag{33}$$

$$c_{V-2} = \frac{16\pi k_B{}^2}{c^3h}T\left\{\left[P_1(x_1)-P_1(x_2)\right]+\frac{1}{2}\left(x_1^2\mathrm{Li}_0(e^{-x_1})-x_2^2\mathrm{Li}_0(e^{-x_2})\right)\right\}. \tag{34}$$

d) Pressure $p$

$$p = \tilde{a}p_0 + \tilde{b}p_{-1} + \tilde{c}p_{-2}, \tag{35}$$

where

$$p_0 = \frac{16\pi k_B{}^4}{c^3h^3}T^4\left\{\left[P_3(x_1)-P_3(x_2)\right]-\frac{1}{6}\left(x_1^3\mathrm{Li}_1(e^{-x_1})-x_2^3\mathrm{Li}_1(e^{-x_2})\right)\right\}; \tag{36}$$

$$p_{-1} = \frac{8\pi k_B{}^3}{c^3h^2}T^3\left\{\left[P_2(x_1)-P_2(x_2)\right]-\frac{1}{2}\left(x_1^2\mathrm{Li}_1(e^{-x_1})-x_2^2\mathrm{Li}_1(e^{-x_2})\right)\right\}; \tag{37}$$

$$p_{-2} = \frac{8\pi k_B{}^2}{c^3h}T^2\left\{\left[P_1(x_1)-P_1(x_2)\right]-\left(x_1\mathrm{Li}_1(e^{-x_1})-x_2\mathrm{Li}_1(e^{-x_2})\right)\right\}. \tag{38}$$

The internal energy density and enthalpy density can be obtained using the following relationships:

1. Internal energy density $u$

$$u(x_1, x_2, T) = f(x_1, x_2, T) + Ts(x_1, x_2, T); \tag{39}$$

2. Enthalpy density *h*

$$h(x_1, x_2, T) = u(x_1, x_2, T) + p(x_1, x_2, T) . \quad (40)$$

The functions *f*, *s*, and *p* are determined by Eq. 23, Eq. 27, and Eq. 35.

The Gibbs free energy density *g*, by definition, is $g(x_1, x_2, T) = h(x_1, x_2, T) - Ts(x_1, x_2, T)$, thus

$$g(x_1, x_2, T) = 0 \ . \quad (41)$$

The chemical potential density is determined as $\mu = \left(\frac{\partial g}{\partial n}\right)_{T,V}$. Therefore, according to Eq. (41), we obtain

$$\mu(x_1, x_2, T) = 0 \ . \quad (42)$$

## 4. Results

### *4.1. Uranium carbide*

In [10], a gap between the normal spectral emissivity of uranium carbide in the solid and liquid phases at melting/ freezing temperature was observed. Now let us show that there are a slight difference between the thermal radiative and thermodynamic functions of liquid and solid uranium carbide during liquid-to-solid transition.

According to [10], the experimental data on the normal spectral emissivity of UC during the melting and freezing arrests is approximated by Eq. 18. In the frequency range from $0.326\,\text{PHz}$ to $0.545\,\text{PHz}$ the coefficients $\tilde{a}, \tilde{b}$, and $\tilde{c}$ have the following values:

$$\text{Solid UC: } \tilde{a} = 0.75746; \ \tilde{b} = -0.1403 \ \text{PHz}; \ \tilde{c} = 0.01662\,\text{PHz}^2. \quad (43)$$

$$\text{Liquid UC: } \tilde{a} = 0.79998; \ \tilde{b} = -0.2265 \ \text{PHz}; \ \tilde{c} = 0.03508\,\text{PHz}^2. \quad (44)$$

The melting and solidification point of UC is established to be (2780 ± 15) K [10]. Substituting Eq. (43) and Eq. (44) in the expressions for the thermal radiative and thermodynamic functions above, we obtain their calculated values presented in Table 1. As clearly seen, a slight difference between the thermal radiative and thermodynamic functions are observed in the solid and liquid phases. For instance, a gap between $\varepsilon_{t\,\text{solid}}$ and $\varepsilon_{t\,\text{liquid}}$ is $0.055$. As for the number density of photons, we have $n_{t\,\text{solid}} - n_{t\,\text{liquid}} = 0.152 \times 10^{16}\,\text{m}^{-3}$.

*4.2. Plutonium carbide*

Unlike UC, a gap between the normal spectral emissivities of PuC in the solid and liquid phases did not observed [10]. As a result, the coefficients $\tilde{a}, \tilde{b}$, and $\tilde{c}$ in Eq. 18 have the same values in both phases and are equal to:

$$\text{Solid/Liquid PuC: } \tilde{a} = 0.85218;\ \ \tilde{b} = -0.2439\ \text{PHz}\ ;\ \ \tilde{c} = 0.03747\ \text{PHz}^2. \tag{45}$$

According to [10], the peritectic melting temperature of $PuC_{0.84}$ was established to be (1900 ± 20) K. Knowing the values of constants at the melting/freezing point allow to calculate the Stefan-Boltzmann law, total energy density, number density of photons, Helmholtz free energy density, internal energy density, enthalpy density, entropy density, heat capacity at constant volume, pressure, and total emissivity in a finite range of frequencies from $0.326\,\text{PHz}$ to $0.545\,\text{PHz}$. Their values are presented in Table 1 (third column). As seen in the Table 1, the normal total emissivity and number density of photons in solid and liquid states are equal to: a) $\varepsilon_{t\,\text{solid}} = \varepsilon_{t\,\text{liquid}} = 0.470$ ; and b) $n_{t\,\text{solid}} = n_{t\,\text{liquid}} = 6.169 \times 10^{14}\,\text{m}^{-3}$.

In conclusion, we note the following. In Table 1, the thermal radiative and thermodynamic functions of solid and liquid UC and PuC are presented when the thermal radiation is emitted by the heated surface per unit area of the sample under study. Therefore, let's calculate the same properties of uranium and plutonium carbides in a case when the thermal radiation is emitted by the heated surface area *S* of the sample. According to [10], the uranium and plutonium carbides samples are presented as a disk about 1 mm thick and around 10 mm in diameter. Then, the surface area *S* and volume *V* of the sample can be defined using the following formulas:

a) Surface area *S*

$$S = 2\pi \left(\frac{d}{2}\right)^2 + \pi h d = 1.885 \times 10^{-4}\,\text{m}^2; \tag{46}$$

b) A volume *V*

$$V = \pi\, h \left(\frac{d}{2}\right)^2 = 7.854 \times 10^{-8}\,\text{m}^3. \tag{47}$$

Using Eq. 46 and Table 1, the total radiation power (Stefan–Boltzmann's law) of solid and liquid UC and PuC emitted by a surface area $S$ of the sample under study take the following values:

a) Uranium carbide

$$\text{Solid UC:}\quad I^{\text{SB}}_{\text{Solidtotal}}(T) = S\, I^{\text{SB}}_{\text{Solid}}(v_1, v_2, T) = 51.74\ \text{W} \tag{48}$$

$$\text{Liquid UC:}\quad I^{\text{SB}}_{\text{Liquidtotal}}(T) = S\, I^{\text{SB}}_{\text{Liquid}}(v_1, v_2, T) = 46.14\ \text{W}\ . \tag{49}$$

b) Plutonium carbide

$$\text{Solid/Liquid PuC:}\quad I^{\text{SB}}_{\text{Solidtotal}}(T) = S\, I^{\text{SB}}_{\text{Solid}}(v_1, v_2, T) = 2.16\ \text{W}\ . \tag{50}$$

In accordance with Eq. (47) and Table 1, the total numbers of photons $N_{\text{total}}$ emitted by the surface area $S$ of solid and liquid UC and PuC are calculated and have the following values:

a) Uranium carbide

$$\text{Solid UC:}\quad N_{\text{Solidtotal}}(T) = V\, n_{\text{Solid}}(v_1, v_2, T) = 1.10 \times 10^{9} \tag{51}$$

$$\text{Liquid UC:}\quad N_{\text{Liquidtotal}}(T) = V\, n_{\text{Liquid}}(v_1, v_2, T) = 9.85 \times 10^{8}\ . \tag{52}$$

b) Plutonium carbide

$$\text{Solid/Liquid PuC:}\quad N_{\text{Solid/Liquidtotal}}(T) = V\, n_{\text{Solid/Liquid}}(v_1, v_2, T) = 4.85 \times 10^{7}\ . \tag{53}$$

In Table 2, the calculated values of the thermal radiative and thermodynamic functions of solid and liquid uranium and plutonium carbides emitted by the heated surface area $S$ of the sample are presented in the in a finite range of frequencies from $0.326\,\text{PHz}$ to $0.545\,\text{PHz}$ at their melting/freezing temperatures. As it can clearly seen, a slight differences between these function for UC in the solid and liquid states are observed. In the case of plutonium carbide, a gap between the thermal radiative and thermodynamic functions is not established.

## 5. Discussion

The current results on uranium carbide confirm the existence of the gaps between the thermal radiative and thermodynamic functions during liquid/ solid phase transition in the visible-near infrared range. For example, a gap of approximately 0.05 between the normal total emissivity of uranium carbide in the liquid and solid states is obtained. For other functions, see Table 1 and Table 2. Unlike UC, the thermal radiative and thermodynamic functions of PuC is insensitive to the phase change on melting.

Now let us compare the obtained results with previously reported data. In [6], based on the experimental data' the recommended value for the normal total emissivity of uranium and plutonium carbides in the temperature range from $T$ = 1250 K to $T$ = 1980 is proposed. This emissivity value is as follows: $\varepsilon_t = 0.42 \pm 0.02$. As it was pointed out in [6] that the value of emissivity may be too low. Note that the melting temperature of plutonium carbide $T$ = 1900 ± 20 K is in this temperature range. Then, according to Table 1, the calculated value of normal total emissivity of PuC is equal to $\varepsilon_t = 0.47$. As seen, a good agreement with experimental data is obtained.

In [6], based on experimental data, the temperature dependence of the mean value of the normal total emissivity of UC with average error 40% was presented analytically in the temperature range between $T$ = 699.8 K and $T$ = 2922.04 K. At the UC melting temperature $T$ = 2780 K, the mean value is equal to $\varepsilon_t$ = 0.34 ± 0.136. According to Table 1, the calculated values of the normal total emissivity of solid and liquid uranium carbide are: a) $\varepsilon_t$ =0.454 - liquid state; and b) $\varepsilon_t$ =0.509 - solid state. As seen, the calculated value $\varepsilon_t$ =0.454 for liquid uranium carbide fell within a narrow band from $\varepsilon_t$ = 0.204 to $\varepsilon_t$ = 0.476. This fact confirms that at the temperature 2780 K melting occurs and a liquid phase exists.

Now let us propose a method to determine the thermal conductivity of carbides at the melting point. It is essential to note that the value of the thermal conductivity of UC at the melting temperature can be obtained by extrapolation high-temperature data to the melting point [5-7]. The exact value of the thermal conductivity in the melting point can't be found in the literature. Now let us consider a method for obtaining the exact value of the thermal conductivity at the melting temperature. This method is based on the Hagen-Rubens relation for the normal total emissivity $\varepsilon_t$ [22] and Wiedemann–Franz law [23]. In the far-infrared region from $5\mathrm{m\mu}$ to $10\mathrm{m\mu}$, the normal total emissivity is presented by the Hagen-Rubens relation [22]

$$\varepsilon_t = 5.78 \left(\frac{T}{\sigma}\right)^{\frac{1}{2}} - 17.8 \left(\frac{T}{\sigma}\right) + 584 \left(\frac{T}{\sigma}\right)^{\frac{3}{2}}, \tag{50}$$

where $\sigma$ is the electrical conductivity in $\Omega^{-1}\mathrm{m}^{-1}$. T is the temperature in K.

The Wiedemann–Franz law is presented by the formula [23].

$$\frac{k}{\sigma} = LT\ , \tag{51}$$

where $k$ is the thermal conductivity and the constant $L$, known as the Lorenz number, is equal to $L = 2.44\times10^{-8}$ W Ω $K^{-2}$ . According to Eq. 51, the electrical conductivity can be written as follows:

$$\sigma = \frac{k}{LT} \tag{52}$$

Substituting Eq 52. in Eq. 50, the expression for the normal total emissivity takes the form,

$$\varepsilon_t = 5.78\ x - 17.8\ x^2 + 584\ x^3. \tag{53}$$

Here $x = \left(\frac{LT^2}{k}\right)^{\frac{1}{2}}$. Finally, the thermal conductivity can be determined as

$$k = \frac{LT^2}{x^2} \tag{54}$$

As seen from Eq. 53 that in determining the thermal conductivity, precise measurements of the normal total emissivity are necessary.

Now let us consider an example of using this method. Assume that the measurement value of $\varepsilon_t$ to be equal to 0.6 in the far-infrared range from $5\mathrm{m}\mu$ to $10\mathrm{m}\mu$ at the melting temperature $T$ =2780 K. Then, solving Eq. 53, we obtain $x \approx 0.0765$. Substituting this value in Eq. 54, the value of the thermal conductivity is equal to $k = 32.22\frac{\mathrm{W}}{\mathrm{m\,K}}$ .

In conclusion, it is important to note that the proposed method allows to obtain the difference between the thermal conductivity in the liquid and solid states at the melting/freezing temperature. This data can be important for understanding the behavior of electrons in carbides on melting, because the difference in the thermal conductivity is related to that between the electronic structures of the liquid and solid phases.

## 6. Conclusion

Thermal radiative and thermodynamic functions of solid and liquid uranium and plutonium carbides at their melting points have been calculated in the visible-near infrared region. The main results of the current study can be summarized as follows.

1. The normal total emissivity of UC is changed during liquid-to-solid transition.

2. The normal total emissivity of UC in the liquid state at the melting point is lower than those of UC in solid state at melting temperature over the measured wavelength range. The gap calculated in the VIS-IR range is about 11%.
3. The normal total emissivity of liquid uranium carbide is equal to $\varepsilon_t = 0.454$ and, unlike solid UC, fell within a narrow experimental band from $\varepsilon_t = 0.204$ to $\varepsilon_t = 0.476$.
4. The total radiation power per unit area of solid UC is slightly larger than those of the liquid UC in the visible - near infrared regions. It means that UC in the solid state emits more radiation than UC in liquid state at the melting temperature. The gap between the total radiation power per unit area of solid and liquid UC is $2.970 \times 10^4$ W m$^{-2}$.
5. The gaps between the thermodynamic functions of UC exist, and their values in the solid state is slightly larger than those of the liquid UC at the melting point. For example, a gap between the entropy density of solid and liquid UC is $1.44 \times 10^{-7}$ J m$^{-3}$ K$^{-1}$.
6. The normal total emissivity of PuC, unlike UC, does not change during liquid-to-solid transition in the frequency range from $0.326$ PHz to $0.545$ PHz.
7. The normal total emissivity at the melting temperature (1900 ± 20) K is 0.47, and in good agreement with previously reported experimental data.
8. The thermodynamic functions of solid and liquid PuC do not sensitive to the liquid-solid phase transition in the $0.326$ PHz - $0.545$ PHz frequency range.
9. A new method to determine the thermal conductivity of metals and carbides at the melting points is proposed.

In conclusion, it should be noted that the general expressions for the thermal radiative and thermodynamic properties obtained in this paper can be applied to the study of the following materials at the melting/freezing temperatures: a) noble metals such as Cu, Ag, and Au [19,20]; and b) iron, cobalt, and nickel metals [21].

These and other topics will be points of discussion in subsequent publications.

**Acknowledgments**

The authors cordially thank Professor L.A Bulavin and Professor N.P. Malomuzh for fruitful discussions.

| Quantity | **Uranium Carbide:** *Solid Phase* | **Uranium Carbide:** *Liquid Phase* | **Plutonium Carbide:** *Solid/Liquid Phases* |
|---|---|---|---|
| $I(\nu_1, \nu_2, T)$ $[\mathrm{J\ m^{-3}}]$ | $3.663\times10^{-3}$ | $3.266\times10^{-3}$ | $1.528\times10^{-4}$ |
| $I^{\mathrm{SB}}\ (\nu_1, \nu_2, T)$ $[\mathrm{W\ m^{-2}}]$ | $2.745\times10^{5}$ | $2.448\times10^{5}$ | $1.145\times10^{4}$ |
| $\varepsilon$ | 0.509 | 0.454 | 0.470 |
| $f$ $[\mathrm{J\ m^{-3}}]$ | $-8.065\times10^{-4}$ | $-8.518\times10^{-4}$ | $-2.940\times10^{-5}$ |
| $s$ $[\mathrm{J\ m^{-3}\ K^{-1}}]$ | $1.871\times10^{-6}$ | $1.727\times10^{-6}$ | $1.220\times10^{-7}$ |
| $p$ $[\mathrm{J\ m^{-3}}]$ | $8.065\times10^{-4}$ | $8.518\times10^{-4}$ | $2.940\times10^{-5}$ |
| $c_V$ $[\mathrm{J\ m^{-3}\ K^{-1}}]$ | $4.942\times10^{-6}$ | $4.375\times10^{-6}$ | $2.873\times10^{-7}$ |
| $u$ $[\mathrm{J\ m^{-3}}]$ | $4.395\times10^{-3}$ | $3.949\times10^{-3}$ | $2.122\times10^{-4}$ |
| $h$ $[\mathrm{J\ m^{-3}}]$ | $5.204\times10^{-3}$ | $4.801\times10^{-3}$ | $2.416\times10^{-4}$ |
| $n$ $[\mathrm{m^{-3}}]$ | $1.406\times10^{16}$ | $1.254\times10^{16}$ | $6.169\times10^{14}$ |

**Table 1** Calculated values of the thermal radiative and thermodynamic functions of thermal radiation of solid and liquid zirconium carbide emitted by a heated surface per unit area of the sample in a finite range of frequencies $0.326\,\mathrm{PHz} \le \nu \le 0.545\,\mathrm{PHz}$ at the eutectic temperatures: a) $T$ = 2780 K – uranium carbide; b) $T$ = 1900 K – plutonium carbide.

| **Quantity** | **Uranium Carbide:** *Solid Phase* | **Uranium Carbide:** *Liquid Phase* | **Plutonium Carbide** *Solid/Liquid Phases* |
|---|---|---|---|
| $I_{\text{total}}(\nu_1,\nu_2,T)$ $[\text{J}]$ | $2.88\times10^{-10}$ | $2.56\times10^{-10}$ | $1.20\times10^{-11}$ |
| $I^{\text{SB}}_{\text{total}}(\nu_1,\nu_2,T)$ $[\text{W}]$ | 51.74 | 46.14 | 2.16 |
| $F_{\text{total}}$ $[\text{J}]$ | $-6.33\times10^{-11}$ | $-6.69\times10^{-11}$ | $-2.31\times10^{-12}$ |
| $S_{\text{total}}$ $[\text{J K}^{-1}]$ | $1.47\times10^{-13}$ | $1.36\times10^{-13}$ | $9.58\times10^{-15}$ |
| $P_{\text{total}}$ $[\text{J}]$ | $6.33\times10^{-11}$ | $6.69\times10^{-11}$ | $2.31\times10^{-12}$ |
| $C_{V\,\text{total}}$ $[\text{J K}^{-1}]$ | $3.88\times10^{-13}$ | $3.44\times10^{-13}$ | $2.26\times10^{-14}$ |
| $U$ $[\text{J}]$ | $3.45\times10^{-10}$ | $3.11\times10^{-10}$ | $1.59\times10^{-11}$ |
| $H$ $[\text{J}]$ | $4.08\times10^{-10}$ | $3.78\times10^{-10}$ | $1.82\times10^{-11}$ |
| $N_{\text{total}}$ | $1.10\times10^{9}$ | $9.85\times10^{8}$ | $4.85\times10^{7}$ |

**Table 2** Calculated values of the thermal radiative and thermodynamic functions of thermal radiation of solid and liquid UC and PuC emitted by a heated surface area $S$ of the sample in the finite frequency range $0.326\,\text{PHz} \le \nu \le 0.545\,\text{PHz}$. The surface area and the volume of the sample are: a) $S = 1.885\times10^{-4}\,\text{m}^2$; b) $V = 7.854\times10^{-8}\,\text{m}^3$.